\documentclass[12pt,a4paper,fleqn,notitlepage,onecolumn]{article}

\usepackage[english]{babel}
\usepackage[ansinew]{inputenc}
\usepackage{amssymb,amsmath}
\usepackage{graphicx} 
\usepackage{fancyhdr} 
\usepackage{natbib} 
\usepackage{wrapfig}
\usepackage{geometry}
\usepackage{ulem}
\usepackage{color}
\newcommand{\prob}{\textrm{prob}}

\newcommand{\mat}{\boldsymbol}

\title{Error estimation in astronomy: A guide}

\author{Ren\'e Andrae\\
\footnotesize{Max-Planck-Institut f\"ur Astronomie, K\"onigstuhl 17, 69117 Heidelberg, Germany}\\ \footnotesize{e-mail: andrae@mpia-hd.mpg.de}}
\date{}

\begin{document}

\maketitle

\begin{center}
\begin{minipage}{15cm}
\small{Estimating errors is a crucial part of any scientific analysis. Whenever a parameter is estimated (model-based or not), an error estimate is necessary. Any parameter estimate that is given without an error estimate is meaningless. Nevertheless, many (undergraduate or graduate) students have to teach such methods for error estimation to themselves when working scientifically for the first time. This manuscript presents an easy-to-understand overview of different methods for error estimation that are applicable to both model-based and model-independent parameter estimates. These methods are not discussed in detail, but their basics are briefly outlined and their assumptions carefully noted. In particular, the methods for error estimation discussed are grid search, varying $\chi^2$, the Fisher matrix, Monte-Carlo methods, error propagation, data resampling, and bootstrapping. Finally, a method is outlined how to propagate measurement errors through complex data-reduction pipelines.}
\end{minipage}
\end{center}

\section{Introduction}

This manuscript is intended as a guide to error estimation for parameter estimates in astronomy. I try to explain several different approaches to this problem, where the emphasis is on highlighting the diversity of approaches and their individual assumptions. Making those assumptions explicitly clear is one of the major objectives, because using a certain method in a situation where its assumptions are \textit{not} satisfied will result in incorrect error estimates. As this manuscript is just an overview, the list of methods presented is by no means complete.

A typical task in scientific research is to make measurements of certain data and then to draw inferences from them. Usually, the inference is not drawn directly from the data but rather from one or more \textit{parameters} that are estimated from the data. Here are two examples:
\begin{itemize}
\item Apparent magnitude of stars or galaxies. Based on a photometric image, we need to estimate the parameter ``flux'' of the desired object, before we can infer its apparent magnitude.
\item Radial velocity of stars. First, we need to take a spectrum of the star and identify appropriate emission/absorption lines. We can then estimate the parameter ``radial velocity'' from fitting these spectral lines.
\end{itemize}
Whenever such parameter estimates are involved, it is also crucial to estimate the error of the resulting parameter.

What does a parameter estimate and its error actually signify? More simply, what is the meaning of an expression such as $4.3\pm 0.7$? This question will be answered in detail in Sects. \ref{sect:parameter_estimate} and \ref{sect:confidence_intervals}, but we want to give a preliminary answer here for the sake of motivation. The crucial point is that the true result of some parameter estimate is \textit{not} something like $4.3\pm 0.7$, but rather a \textit{probability distribution} for all possible values of this parameter. An expression like $4.3\pm 0.7$ is nothing more than an attempt to encode the information contained in this probability distribution in a more simple way, where the details of this ``encoding'' are given by some general standards (cf. Sect. \ref{sect:confidence_intervals}). Put simply, the value 4.3 signifies the maximum of the probability distribution (the most likely value), whereas the ``error'' 0.7 signifies the width of the distribution. Hence, the value 4.3 alone contains insufficient information, since it does not enable us to reconstruct the probability distribution (the true result). More drastically: \textit{A parameter value without a corresponding error estimate is meaningless.} Therefore, error estimation is equally as important an ingredient in scientific work as parameter estimation itself. Unfortunately, a profound and compulsory statistical eduction is missing in many university curricula. Consequently, when (undergraduate or graduate) students are faced with these problems for the first time during their research, they need to teach it to themselves. This is often not very efficient and usually the student focuses on a certain method but does not gain a broader overview. The author's motivation was to support this process by providing such an overview.

Where do uncertainties stem from? Of course, an important source of uncertainties is the measured dataset itself, but models can also give rise to uncertainties. Some general origins of uncertainties are:
\begin{itemize}
\item Random errors during the measurement process.
\item Systematic errors during the measurement process.
\item Systematic errors introduced by a model.
\end{itemize}
We obviously have to differentiate between random and systematic errors, i.e., between variance/scatter and bias/offset. Systematic errors are usually very hard to identify and to correct for. However, this is \textit{not} part of this manuscript, since individual solutions strongly depend on the specific problem. Here, different methods of estimating random errors (variance/scatter) are considered, i.e., those quantities that determine the size of error bars or, more generally, error contours. Error estimation for parameter estimation only is described, whereas error estimation for classification problems is not discussed.

This manuscript will not be submitted to any journal for two reasons: First, its content is not genuinely new but a compilation of existing methods. Second, its subject is statistical methodology rather than astronomy. Any comments that may improve this manuscript are explicitly welcome.

\section{Preliminaries}

Before diving into the different methods for error estimation, some preliminaries should be briefly discussed, firstly, the terminology used throughout this manuscript and, secondly, errors of data. Thirdly, the basics of parameter estimation are briefly explained, including the introduction of the concept of a likelihood function. Fourthly, the central-limit theorem is discussed. Finally, the concept of confidence intervals, which are the desired error estimates, is introduced.

\subsection{Terminology}

This manuscript is about ``error estimation for parameter estimates''. The first step is usually to measure some data and also to \textit{measure} its error or uncertainty (Sect. \ref{sect:data_errors}). Given this \textit{measurement}, the task is then to \textit{estimate} some parameter (Sect. \ref{sect:parameter_estimate}). The estimate of the parameter $\theta$ is denoted by a hat, $\hat\theta$, which is common practice in statistics. Here, I want to introduce the concept of a qualitative difference between ``measuring'' and ``estimating'': Measurements are outcomes of a real experiment. Conversely, estimates are inferences from measurements, i.e., they are \textit{not} directly related to experiments. Although this difference is not of vital importance, both terms are rigorously differentiated throughout this manuscript in order to make clear what is being referred to.

Another issue of terminology concerns the words ``error'' and ``uncertainty''. As mentioned in the introduction, systematic errors are not considered here, and hence both terms may be used more or less synonymously.\footnote{\citet{Hogg2010} argue that the word ``error'' is not a good terminology anyway and that it should be replaced by ``uncertainty''. However, their argument assumes that systematic errors (biases) can always be corrected.} There is also a third word of the same family, namely ``noise'', which could also be used synonymously. However, personally I would use the word ``noise'' only in the context of measurements, whereas in the context of parameter estimates the word ``uncertainty'' appears to be most natural.

\subsection{Errors of data\label{sect:data_errors}}

Real data is always subject to noise, i.e., there is always some uncertainty.\footnote{One can think of real data as being one of many different ``noise realisations'' of the truth, but not being the truth itself.} The precise origin of the noise, e.g., read-out noise, sky noise, etc., is not relevant for this manuscript. Most methods of error estimation for parameters require knowledge about the \textit{error distribution} (or noise distribution) of the measured data. Random scatter (variance, noise) determines the width of the error distribution, whereas the mean of the error distribution is determined by the physical signal which may be shifted by a systematic error (bias). The easiest way to \textit{measure} the data's error distribution is to repeat an identical measurement process and to monitor the distribution of the results. However, this is often infeasible, e.g., because a measured event happens rarely or because the measurement process itself is expensive (in time, money, computational effort, etc.).

\begin{figure}
  \begin{minipage}[t]{0.45\textwidth}\centering
\includegraphics[width=8cm]{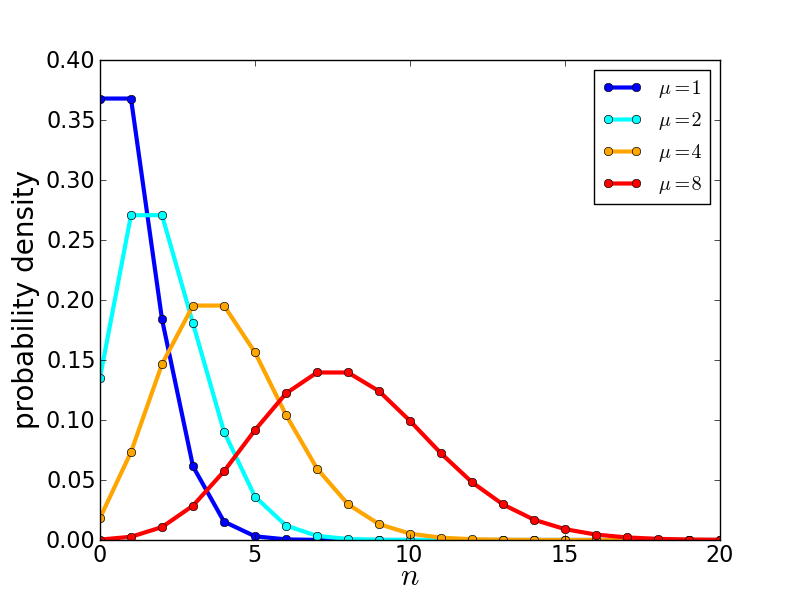}
\caption{Examples of Poisson distributions with $\mu=1$, $\mu=2$, $\mu=4$, and $\mu=8$.}
\label{fig:examples_Poisson_distribution}
  \end{minipage}
\hspace{0.05\textwidth}
  \begin{minipage}[t]{0.45\textwidth}\centering
\includegraphics[width=8cm]{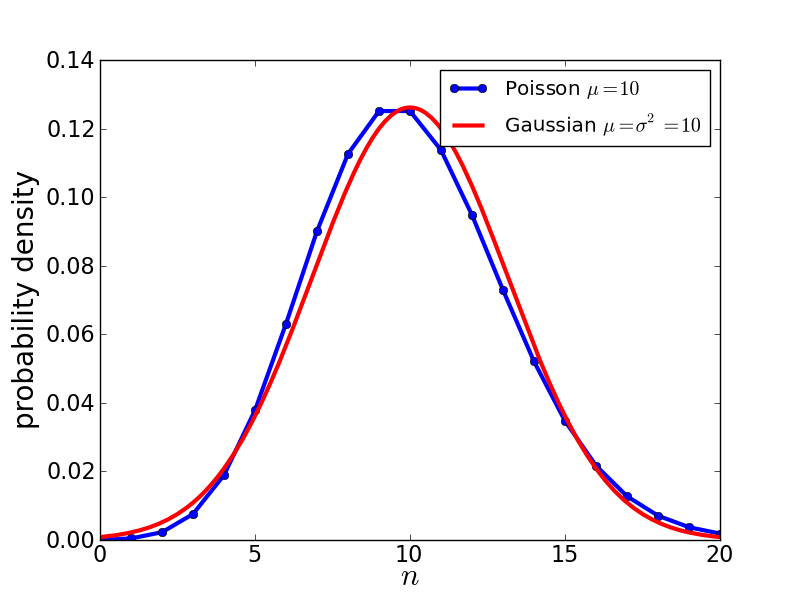}
\caption{Poisson distribution with $\mu=10$ and Gaussian distribution with $\sigma^2=\mu=10$. The Gaussian is a very good approximation to the Poisson distribution.}
\label{fig:comparing_Gaussian_Poisson}
  \end{minipage}
\end{figure}

Fortunately, for many measurement processes in astronomy the mathematical type of the error distribution is known from \textit{physical} arguments, e.g., whenever spectroscopy or photometry are carried out. In this case, the physical measurement process is counting photons in a certain pixel. This photon counting is usually assumed to be a Poisson process, i.e., it follows the \textit{Poisson distribution}. Given a mean value of $\mu$ photons in a certain pixel, the probability of measuring $n$ photons in this pixel is given by
\begin{equation}\label{eq:poission_dist}
\prob(n|\mu) = \frac{\mu^n}{n!}e^{-\mu} \,\textrm{.}
\end{equation}
Figure \ref{fig:examples_Poisson_distribution} shows examples of Poisson distributions with different mean values. In practice, the mean $\mu$ is usually unknown and the task is to estimate the value of $\mu$ given $n$. We emphasise that the mean of the Poisson distribution does \textit{not} coincide with the maximum position (the \textit{mode}), since the Poisson distribution is not symmetric. The fact that the Poisson distribution is defined for \textit{positive} integers only already implies that it cannot be symmetric.

Modern instruments have high sensitivities and it is also common practice (if possible) to choose the exposure time in the measurement such that the number of photon counts per pixel is large. The obvious reason to do this is to improve the signal-to-noise ratio. However, there is also another benefit: If the number of photon counts is large, the expected value of $\mu$ will also be large. In the case of large $\mu$, the Poisson distribution approaches the Gaussian distribution,
\begin{equation}\label{eq:gaussian_dist}
\prob(n|\mu,\sigma) = \frac{1}{\sqrt{2\pi\sigma^2}}\exp\left[-\frac{1}{2}\frac{(n-\mu)^2}{\sigma^2}\right] \,\textrm{,}
\end{equation}
which has much more convenient analytical properties than the Poisson distribution, as we shall see in Sections \ref{sect:parameter_estimate}, \ref{sect:varying_chi2}, and \ref{sect:fisher_matrix}. Figure \ref{fig:comparing_Gaussian_Poisson} shows a Poisson distribution with mean $\mu=10$ and a Gaussian with mean $\mu=10$ and variance $\sigma^2=\mu=10$. Evidently, for a mean of only ten photon counts per pixel, the actual Poisson distribution can be nicely approximated by a Gaussian already. This is usually a valid approximation in the optical regime and at larger wavelengths, whereas in the high-energy regime (UV, X-ray, gamma) it is not unheard of that there are less than ten photon counts per pixel.

\subsection{Parameter estimation\label{sect:parameter_estimate}}

I now briefly discuss the concept of parameter estimation. Here, the focus is on model-based estimates (e.g.\ estimating a mean), because they all have a common concept, whereas model-independent parameter estimation (e.g.\ estimating image flux) is a very diverse field and usually self-explanatory.

The common feature of model-based parameter estimation is that it is usually an optimisation (or ``fitting'') problem. The task at hand is to optimise a certain score function, e.g., to minimise residuals or -- more technically -- to maximise the \textit{likelihood function}\footnote{Bayesians prefer to maximise the posterior probability instead. I will restrict the discussion on likelihood functions alone, since Bayesians should be happy with that assuming flat priors, whereas discussing posterior probabilities would probably enrage some frequentists.}. For a given set of measured data $D$ and a given model $M$ with parameters $\theta$, the likelihood function is defined via the \textit{noise model},
\begin{equation}\label{eq:def:likelihood_function}
\mathcal L(D;M,\theta) = \prob(D|M,\theta) \,\textrm{.}
\end{equation}
In words, the likelihood is the probability of measuring the data $D$ we have, given the model $M$ with parameters $\theta$.\footnote{Quoting \citet{Heavens2009}: If you are confused by the conditional probabilities $\prob(A|B)$ of $A$ given $B$ and $\prob(B|A)$ of $B$ given $A$ consider if $A$=pregnant and $B$=female. Then $\prob(A|B)$ is a few percent whereas $\prob(B|A)$ is unity.} Maximum-likelihood estimation then means that we choose the parameter values such that the data we \textit{did} measure were the most likely measurement outcome.\footnote{Note the subtle fact that this is actually \textit{not} the question we are asking. We actually want to know what are the most likely parameters given the data and model. Hence, we should actually maximise $\prob(\theta|M,D)$ rather than $\prob(D|M,\theta)$. However, this would lead us to Bayes' theorem and the issue of priors \citep[e.g.\ see][]{Barlow1993} which are the cause of the long-ongoing Bayesian vs. frequentist discussion in the scientific community.} The set of parameters that maximise the likelihood function is called the \textit{maximum-likelihood estimate} and it is usually denoted by $\hat\theta$.

A somewhat philosopical note: Actually, the likelihood function as given by Eq. (\ref{eq:def:likelihood_function}) is what every parameter estimation is aiming for. This function, $\mathcal L(D;M,\theta)$, contains all important information about the data and the model, a theorem which is called \textit{likelihood principle}. However, Eq.\ (\ref{eq:def:likelihood_function}) is just an abstract definition and even a more specific example (e.g.\ Eqs.\ (\ref{eq:prop_xn_gauss}) and (\ref{eq:lik_gauss})) usually does not provide more insight. Therefore, one has to extract the information from Eq. (\ref{eq:def:likelihood_function}) in some way. If the model under consideration has only one or two model parameters, it is possible to plot the likelihood function directly (e.g.\ Fig.\ \ref{fig:example_brute_force_grid}), without involving any optimisation procedure. Although such a plot is actually the final result of the parameter-estimation process, people (including myself) are usually happier giving ``numbers''. Moreover, if a model has more than two parameters, the likelihood function cannot be plotted anymore. Hence, the standard practise of encoding the information contained in the likelihood function is by identifying the point in parameter space where the likelihood function takes its maximum (the maximum-likelihood estimate) plus inferring the ``width'' of the function at its maximum (the uncertainty). If nothing contrary is said, these two quantities usually signify the mean value and the standard deviation of a Gaussian. Consequently, if both values are provided, one can reconstruct the full likelihood function.

\begin{table}\centering
\begin{tabular}{cc|cc|cc|cc|cc}
\hline\hline
value $x_n$ & error & value $x_n$ & error & value $x_n$ & error & value $x_n$ & error & value $x_n$ & error \\
\hline
 7 & 3.18 & 12 & 3.18 & 12 & 3.08 & 11 & 2.87 &  8 & 3.32 \\
10 & 3.45 &  9 & 3.14 & 11 & 3.41 &  7 & 3.07 & 11 & 2.99 \\
11 & 2.92 & 12 & 3.12 & 13 & 3.32 &  7 & 2.89 &  9 & 3.08 \\
10 & 3.14 & 13 & 3.03 &  9 & 3.44 & 12 & 3.06 &  9 & 3.18 \\
 8 & 3.43 & 10 & 3.12 & 10 & 3.31 & 10 & 2.93 &  9 & 3.40 \\
14 & 2.85 & 11 & 3.07 & 12 & 3.21 &  6 & 2.90 &  9 & 3.01 \\
\hline
\end{tabular}
\caption{Data sample used as a standard example for all methods. All data points $x_n$ are sampled from a Poisson distribution with mean $\mu=10$ (cf. Fig. \ref{fig:comparing_Gaussian_Poisson}). The columns entitled ``error'' give the Gaussian standard deviations $\sigma_n$ for each data point $x_n$ for the cases where the error distribution is assumed to be Gaussian.}
\label{tab:standard_data_set}
\end{table}

In order to ``give some flesh'' to the rather abstract concept of a likelihood function, two simple examples of parameter estimation are now discussed. This allows us to see this concept and the Poisson and Gaussian distributions ``in action''. Table \ref{tab:standard_data_set} also introduces the data sample that will be used to demonstrate every method that is discussed using actual numbers.

\subsubsection{Example 1: Estimating the mean of a Gaussian distribution\label{sect:example_1}}

The first example is to estimate the mean $\mu$ of a sample of $N$ measured data points $D=\{x_1,x_2,\ldots,x_N\}$ that are all real values. The \textit{assumption} is that all data points have a Gaussian error distribution, i.e.,
\begin{equation}\label{eq:prop_xn_gauss}
\prob(x_n|\mu,\sigma_n) = \frac{1}{\sqrt{2\pi\sigma_n^2}}\exp\left[-\frac{1}{2}\frac{(x_n-\mu)^2}{\sigma_n^2}\right]  \,\textrm{,}
\end{equation}
where we assume that each measurement $x_n$ has its own standard deviation $\sigma_n$. The probability of all $N$ measurements $D=\{x_1,x_2,\ldots,x_N\}$ -- the likelihood function -- is (in the case of uncorrelated measurements) just the product of the probabilities of the individual measurements, i.e.,
\begin{equation}\label{eq:lik_gauss}
\mathcal L(D;\mu) = \prod_{n=1}^N\prob(x_n|\mu,\sigma_n)  \,\textrm{.}
\end{equation}
There are two reasons for maximising $\log\mathcal L$ instead of $\mathcal L$. First, $\log\mathcal L$ sometimes takes a much more convenient mathematical form, enabling us to solve the maximisation problem analytically, as we shall see immediately. Second, $\mathcal L$ is a product of $N$ potentially small numbers. If $N$ is large this can cause a numerical underflow in the computer. As the logarithm is a strictly monotonic function, the maxima of $\mathcal L$ and $\log\mathcal L$ will be identical. The \textit{logarithmic likelihood function} is given by
\begin{equation}
\log\mathcal L(D;\mu) = \sum_{n=1}^N\log\prob(x_n|\mu,\sigma_n)  \,\textrm{.}
\end{equation}
Inserting Eq. (\ref{eq:prop_xn_gauss}) yields
\begin{equation}\label{eq:one}
\log\mathcal L(D;\mu) = -\frac{1}{2}\sum_{n=1}^N\frac{(x_n-\mu)^2}{\sigma_n^2} + C  \,\textrm{,}
\end{equation}
where $C$ encompasses all terms that do not depend on $\mu$ and are therefore constants during the maximisation problem. We can now identify the sum,
\begin{equation}\label{eq:def:chi2}
\chi^2 = \sum_{n=1}^N\frac{(x_n-\mu)^2}{\sigma_n^2}  \,\textrm{,}
\end{equation}
such that $\log\mathcal L(D;\mu) = -\frac{1}{2}\chi^2 + C$. In other words, maximising the likelihood function in the case of Gaussian noise is equivalent to minimising $\chi^2$.\footnote{Actually, we should say that minimising $\chi^2$ provides the correct estimator \textit{if and only if} the error distribution of the data is Gaussian. If the error distribution is not Gaussian, then minimising squared residuals may well be \textit{plausible} but it is \textit{not justified}.} In order to estimate $\mu$, we now take the first derivative of Eq. (\ref{eq:one}) or Eq. (\ref{eq:def:chi2}) with respect to $\mu$, set it equal to zero and try to solve the resulting equation for $\mu$. The first derivative of Eq. (\ref{eq:one}) set to zero then reads
\begin{equation}\label{eq:first_derivative}
\frac{d\log\mathcal L(D;\mu)}{d\mu} = \sum_{n=1}^N\frac{(x_n-\mu)}{\sigma_n^2} =0  \,\textrm{,}
\end{equation}
and solving for $\mu$ yields the maximum-likelihood estimate
\begin{equation}\label{eq:ex1_general_estimator}
\hat\mu = \frac{\sum_{n=1}^N\frac{x_n}{\sigma_n^2}}{\sum_{n=1}^N\frac{1}{\sigma_n^2}}  \,\textrm{.}
\end{equation}
This \textit{estimator} is a weighted mean which underweights data points with large measurement errors, i.e., data points that are very uncertain. For the example data set of Table \ref{tab:standard_data_set} we get $\hat\mu\approx 10.09$. This result can be simplified by assuming that all data points have identical standard deviations, i.e., $\sigma_n=\sigma$ for all $x_n$. Our result then reads
\begin{equation}
\hat\mu = \frac{\sum_{n=1}^N x_n}{N}  \,\textrm{,}
\end{equation}
which is simply the arithmetic mean. For the example data set of Table \ref{tab:standard_data_set} we then get $\hat\mu\approx 10.07$. The derivation of the corresponding error estimation of $\hat\mu$ is postponed to Sect. \ref{sect:fisher_matrix}.

\subsubsection{Example 2: Estimating the mean of a Poisson distribution}

The second example is precisely the same task, but now the error distribution of the $N$ measurements $D=\{x_1,x_2,\ldots,x_N\}$ (which are now all integers) should be a Poisson distribution as given by Eq. (\ref{eq:poission_dist}). Again, we estimate the mean $\mu$ by maximising the likelihood function of the data
\begin{equation}
\mathcal L(D;\mu) = \prod_{n=1}^N\prob(x_n|\mu)  \,\textrm{,}
\end{equation}
or rather the logarithmic likelihood function
\begin{equation}
\log\mathcal L(D;\mu) = \sum_{n=1}^N \log\left(\frac{\mu^{x_n}}{x_n!}e^{-\mu}\right) = -N\,\mu + \log(\mu)\sum_{n=1}^N x_n + C  \,\textrm{,}
\end{equation}
where $C$ again summarises all terms that do not depend on $\mu$. Taking the first derivative with respect to $\mu$ and setting it to zero yields
\begin{equation}\label{eq:ex2_first_derivative}
\frac{d\log\mathcal L(D;\mu)}{d\mu} = -N + \frac{1}{\mu}\sum_{n=1}^N x_n = 0  \,\textrm{.}
\end{equation}
Solving for $\mu$ then yields the maximum-likelihood estimate
\begin{equation}\label{eq:poisson_estimator}
\hat\mu = \frac{1}{N}\sum_{n=1}^N x_n  \,\textrm{,}
\end{equation}
which is the arithmetic mean, again. Do not be mistaken by the fact that the result was identical for the Gaussian and the Poisson distribution in this example. In general, trying to estimate a certain quantity assuming different error distributions also results in different estimators.

\subsubsection{Example 3: Estimating a fraction}

\begin{wrapfigure}{right}{8cm}
\includegraphics[width=8cm]{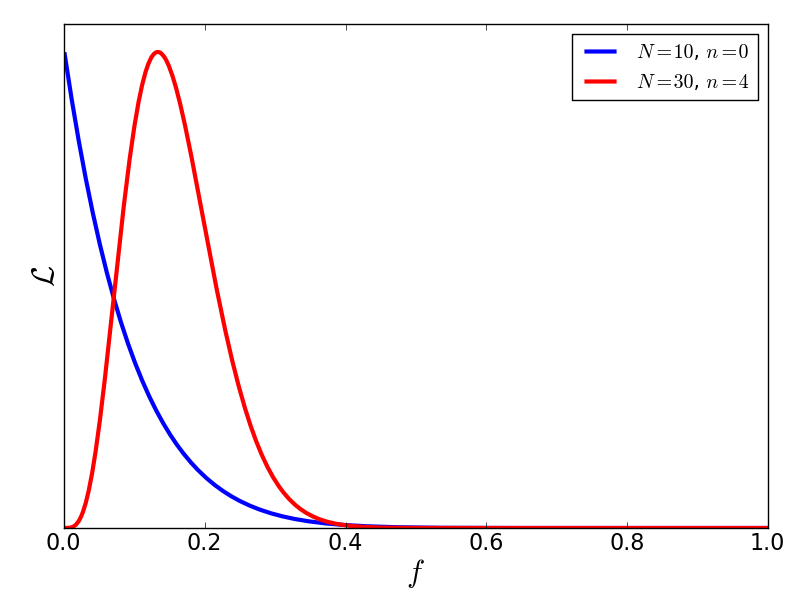}
\caption{Binomial likelihood functions for the fraction $f\in[0,1]$ for example 3. For the sake of visualisation the likelihood functions are \textit{not} normalised, which is why the $y$-axis is unlabelled.}
\label{fig:binomial_distributions}
\end{wrapfigure}

Finally, we also want to consider a third example which will turn out to be less well behaved than the first two. Let us consider the situation that we are given a set of $N$ objects, say galaxies, of which $n$ are in some way different from the other $N-n$ objects, say they are active galaxies. The task is to estimate the fraction $f$ of special objects from these numbers. The natural assumption here is that the likelihood function is a binomial distribution, i.e.,
\begin{equation}
\mathcal L(D;f)=\mathcal L(n,N;f) = \frac{n!}{N!(N-n)!}f^n(1-f)^{N-n} \;\textrm{.}
\end{equation}
The desired fraction is limited to the interval $f\in[0,1]$. What value of $f$ maximises the likelihood of the measured values of $N$ and $n$? Let us first compute the logarithmic likelihood,
\begin{equation}\label{eq:binomial_loglik}
\log\mathcal L(D;f) = n\log f + (N-n)\log(1-f) + C \;\textrm{,}
\end{equation}
where $C$ contains everything that does not depend on $f$. The first derivative of $\log\mathcal L$ w.r.t.\ $f$ reads,
\begin{equation}
\frac{\partial\log\mathcal L(D;f)}{\partial f} = \frac{n}{f} - \frac{N-n}{1-f} \;\textrm{.}
\end{equation}
Equating this to zero and solving for $f$ yields the maximum-likelihood estimator
\begin{equation}
\hat f=\frac{n}{N} \;\textrm{,}
\end{equation}
provided that neither $n$ nor $f$ nor $N$ are zero.\footnote{This result should not come as a suprise. That the fraction is estimated via $n/N$ would have been a \textit{plausible guess}, but mind that we now have seen why it is \textit{justified}.} We will consider this example in two flavours:
\begin{enumerate}
\item $N=10$ and $n=0$, i.e., the given sample is very small and contains no special objects. An error estimate enables us to assess whether this rules out the existence of these special objects.
\item $N=30$ and $n=4$.
\end{enumerate}
Figure \ref{fig:binomial_distributions} shows the binomial likelihood functions for both cases. An example where a binomial distribution shows up in astronomy can, e.g., be found in \citet{Cisternas2010}.

\subsubsection{More general error distributions}

The task of model-based parameter estimation as outlined in the previous subsections can be summarised as follows:
\begin{enumerate}
\item Identify the error distribution of the data and write down the (logarithmic) likelihood function.
\item In order to maximise the (logarithmic) likelihood function, take the first derivative(s) with respect to the desired model parameter(s) and set it equal to zero.
\item Try to solve the resulting (system of) equation(s) for the desired fit parameter(s).
\end{enumerate}
The first two steps are usually not that hard. In the two examples given above, step 3 was also fairly simple because our model was simple (our model was a constant $\mu$) and the error distributions were well-behaved. However, in general it is not possible to perform step 3 analytically for more complex models or other error distributions. In this case the optimisation has to be done numerically. \citet{MacKay2003} provides an overview of the most important distribution functions that may -- in principle -- appear as error distributions.

Nonetheless, we should not be mistaken by the simplicity of the previous examples. Replacing the error distribution measured in some experiment by an analytic distribution function, such as a Poisson or Gaussian, is merely a makeshift. We have to keep in mind that it may happen in practice that for some measured error distribution it may not be possible to find such an analytic parametrisation. In such cases, things get substantially more difficult as we can no longer write down the likelihood function. Even so, we could still work with the whole unparametrised error distribution.

\subsection{Central-limit theorem\label{sect:CLT}}

The central-limit theorem is a key result of statistics and shows up repeatedly throughout this manuscript. In simple words, the central-limit theorem tells us that if certain regularity conditions are met, \textit{any} likelihood function is \textit{asymptotically} Gaussian near its maximum. Let us denote $\log\mathcal L=\log\mathcal L(\vec\theta)$ as a function of the $P$ parameters $\vec\theta=\{\theta_1,\ldots,\theta_P\}$ and Taylor expand it around the maximum $\vec\theta_\textrm{max}$ to second order,
\begin{equation}\label{eq:expanion_L}
\log\mathcal L(\vec\theta) \approx \log\mathcal L(\vec\theta_\textrm{max}) + \frac{1}{2}\left.\frac{\partial^2\log\mathcal L}{\partial\theta_i\partial\theta_j}\right|_{\vec\theta_\textrm{max}}(\theta-\theta_\textrm{max})_i(\theta-\theta_\textrm{max})_j \,\textrm{.}
\end{equation}
\begin{wrapfigure}{left}{8cm}
\includegraphics[width=8cm]{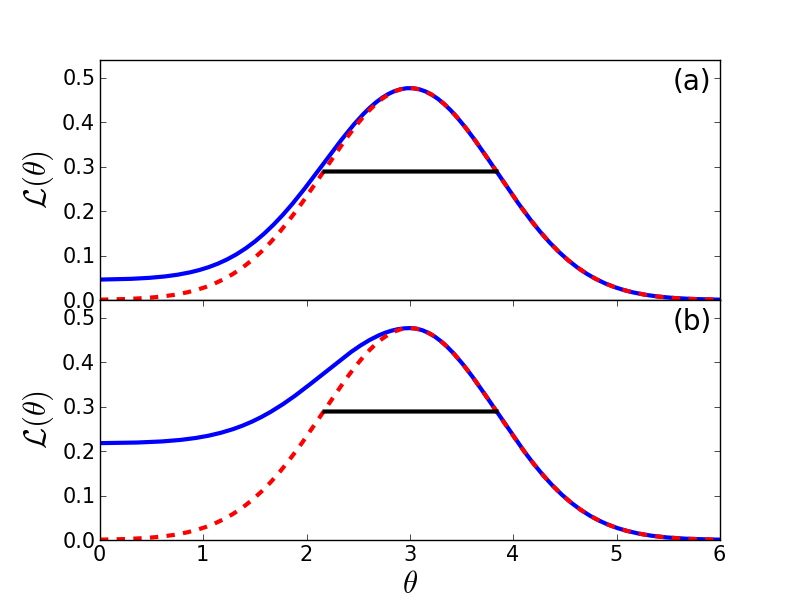}
\caption{Possible failures of the central-limit theorem. This figure shows an example likelihood function $\mathcal L$ as a function of some parameter $\theta$ (solid blue curves), its Gaussian approximation at the maximum (dashed red curves), and the widths of these Gaussians (horizontal solid black lines). In panel (a) the Gaussian approximation is good and the error estimate reliable. In panel (b) the Gaussian approximation is rather poor and the error is substantially underestimated.}
\label{fig:failure_fisher}
\end{wrapfigure}
The linear term vanishes at the maximum, because the first derivative evaluated at $\vec\theta_\textrm{max}$ becomes zero. Consequently, $\log\mathcal L(\vec\theta)$ is approximately a quadratic form in $\vec\theta$ near the optimum, i.e., $\mathcal L(\vec\theta)=e^{\log\mathcal L(\vec\theta)}$ is a Gaussian. The goodness of the Gaussian approximation depends on the specific situation, i.e., on the measured data, its noise, and the model used. The approximation is good, if we are close enough. However, for error estimation we cannot go arbitrarily close to the maximum of the likelihood function. In fact, the Gaussian approximation can be arbitrarily poor in practice. Figure \ref{fig:failure_fisher} shows an example where this is the case. The central-limit theorem may be particularly problematic in case of parameters that are not defined on the interval $(-\infty,\infty)$ of a Gaussian but, e.g., are constrained on the interval $[0,1]$ such as the fraction of example 3. Moreover, there are non-pathologic cases where the likelihood function \textit{never} becomes Gaussian at its maximum. For instance, consider example 3.1 and Fig. \ref{fig:binomial_distributions}, where the likelihood function even does not have a real maximum where its first derivative would be zero. In this particular case, the Taylor expansion of Eq. (\ref{eq:expanion_L}) breaks down, because we cannot compute any derivatives at the ``maximum'' at all. Consequently, whenever invoking the central-limit theorem, one should carefully check the goodness of the Gaussian approximation.

\subsection{Confidence intervals\label{sect:confidence_intervals}}

In order to explain methods for error estimation, confidence intervals have to be considered. We are all familiar with the concept of error bars, error ellipses or -- in general -- error contours. They are different manifestations of confidence intervals. These confidence intervals are inferred from the likelihood function.

\subsubsection{A single parameter}

\begin{figure}
  \begin{minipage}[t]{0.45\textwidth}\centering
    \includegraphics[width=1.0\textwidth]{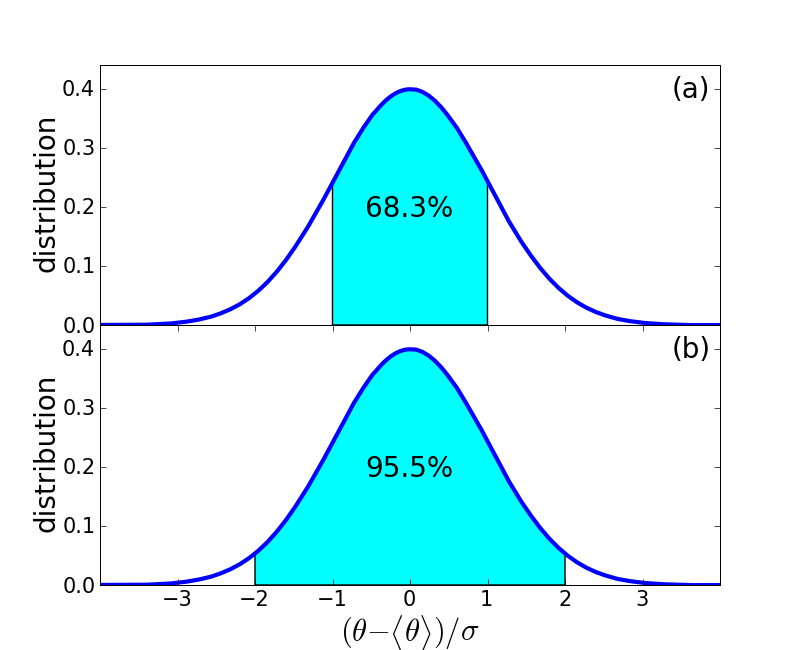}
\caption{Confidence intervals for the Gaussian distribution of mean $\langle\theta\rangle$ and standard deviation $\sigma$. If we draw $N$ values of $\theta$ from a Gaussian distribution, 68.3\% of the values will be inside the interval $[\langle\theta\rangle - \sigma,\langle\theta\rangle + \sigma]$ as shown in panel (a), whereas 95.5\% of the values will be inside the interval $[\langle\theta\rangle - 2\sigma,\langle\theta\rangle + 2\sigma]$ as shown in panel (b).}
\label{fig:confidenceIntervalsGaussian}
  \end{minipage}
\hspace{0.05\textwidth}
  \begin{minipage}[t]{0.45\textwidth}\centering
    \includegraphics[width=1.0\textwidth]{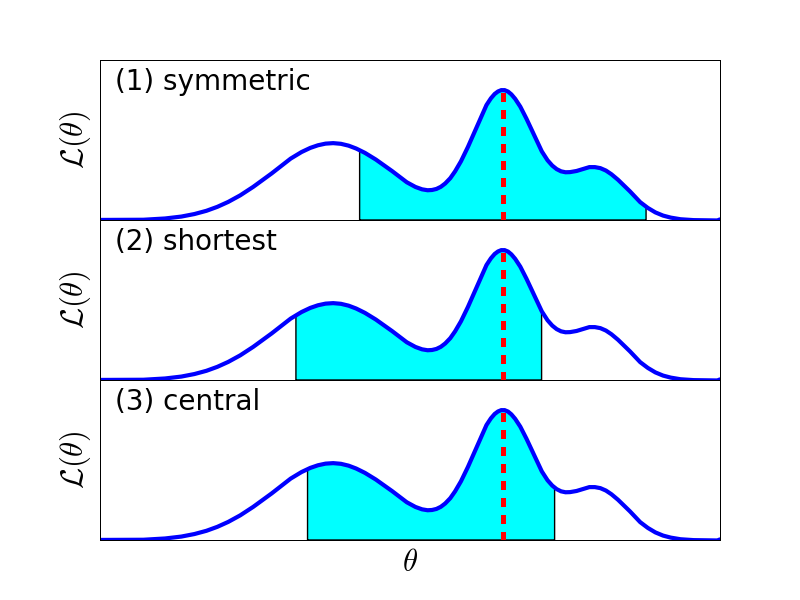}
\caption{Different types of 68.3\% confidence intervals for a multimodal likelihood function. The vertical dashed red line indicates the maximum-likelihood estimate $\hat\theta$. The panels are numbered according to the definitions in the main text.}
\label{fig:confidenceIntervalsGeneral}
  \end{minipage}
\end{figure}

Initially, confidence intervals for a single parameter are considered, in order to keep things simple. Figure \ref{fig:confidenceIntervalsGaussian} shows the most simple example -- confidence intervals of a Gaussian distribution. If we draw a sample value $\theta$ from a Gaussian with mean $\langle\theta\rangle$ and standard deviation $\sigma$, e.g., by trying to estimate $\langle\theta\rangle$ from measured data, the deviation $|\theta - \langle\theta\rangle|$ will be smaller than $1\sigma$ with 68.3\% probability, and it will be smaller than $2\sigma$ with 95.5\% probability, etc. In simple words, if we fit some function to $N$ data points with Gaussian errors, we have to expect that 31.7\% of all data points deviate from this fit by more than one sigma.\footnote{If you are presented fitted data where the fit goes through all $1\sigma$-errorbars, you should definitely be sceptical.}

The Gaussian is an almost trivial example, due to its symmetry around the mean. In general, likelihood functions may not be symmetric and how to define confidence intervals in these cases should be explained. For asymmetric distributions mean and maximum position do not conincide (e.g.\ Poisson distribution). The actual parameter estimate $\hat\theta$ is the \textit{maximum}-likelihood estimate, i.e., it indicates the maximum of the likelihood function, not its mean. We define the confidence interval $\theta_-\leq\hat\theta\leq\theta_+$ for a given distribution function $\prob(\theta)$ via \citep[e.g.][]{Barlow1993}
\begin{equation}\label{eq:def:confidence_interval}
\prob(\theta_-\leq\hat\theta\leq\theta_+) = \int_{\theta_-}^{\theta_+}d\theta\,\prob(\theta) = C \,\textrm{,}
\end{equation}
where usually $C=0.683$ in analogy to the one-sigma-interval of the Gaussian. In practice, the distribution function $\prob(\theta)$ is usually unknown and only given as a histogram of samples of $\theta$. In this case, the integral in Eq. (\ref{eq:def:confidence_interval}) reduces to the fraction of all samples $\theta$ that are between $\theta_-$ and $\theta_+$. However, Eq. (\ref{eq:def:confidence_interval}) does \textit{not} uniquely define the confidence interval, an additional criterion is required. Possible criteria are \citep[e.g.][]{Barlow1993}:
\begin{enumerate}
\item Symmetric interval: $\theta_-$ and $\theta_+$ are symmetric around the parameter estimate, i.e., $\hat\theta-\theta_-=\theta_+ - \hat\theta$.
\item Shortest interval: $\theta_+-\theta_-$ is smallest for all intervals that satisfy Eq. (\ref{eq:def:confidence_interval}).
\item Central interval: The probabilities above and below the interval are equal, i.e., $\int_{-\infty}^{\theta_-}d\theta\,\prob(\theta)=\int^\infty_{\theta_+}d\theta\,\prob(\theta)=(1-C)/2$.
\end{enumerate}
In case of a symmetric distribution, e.g., a Gaussian, all three definitions are indeed equivalent. However, in general they lead to different confidence intervals. Figure \ref{fig:confidenceIntervalsGeneral} shows the 68.3\% confidence intervals\footnote{Note our terminology: We are talking of a ``68.3\% confidence interval'', not of a ``one-sigma interval''.} resulting from the three definitions for an example distribution that could be a likelihood function resulting from a parameter estimate. In practice, there is usually no preference for any of these definitions\footnote{A symmetric confidence interval may not be sensible in case of a highly asymmetric likelihood function. As a nice example, consider the likelihood function of example 3.1 shown in Fig. \ref{fig:binomial_distributions}. Furthermore, the central interval would cause the ``maximum'' at $\hat f=0$ to lie \textit{outside} this confidence interval.}, it should only be made explicitly clear which one is used.

\subsubsection{Two or more parameters}

If we are estimating two or more parameters and are interested in estimating the joint confidence \textit{region}, things become considerably more difficult. This difficulty largely stems from the fact that multiple parameters will usually exhibit mutual correlations. The following discussion largely follows \citet{Barlow1993}.

First, consider the case where the central-limit theorem indeed ensures that some (multidimensional) likelihood function is approximately Gaussian at its maximum position. Such (multivariate) Gaussians, with $P$-dimensional mean vector $\vec\mu$ and $P\times P$ covariance matrix $\mat\Sigma$,
\begin{equation}
\prob(\vec x|\vec\mu,\mat\Sigma) = \frac{1}{\sqrt{(2\pi)^P \det\mat\Sigma}}\exp\left[ -\frac{1}{2}(\vec x-\vec\mu)^T\cdot\mat\Sigma^{-1}\cdot(\vec x-\vec\mu) \right] \,\textrm{,}
\end{equation}
provide ellipsoidal error contours, i.e., they are capable of describing \textit{linear} correlations in the parameters, but not nonlinear correlations such as ``banana-shaped'' error contours. However, even in this simple case, things are complicated. The reason for this is that the one-sigma contour no longer marks a 68.3\% confidence region as in Fig. \ref{fig:confidenceIntervalsGaussian}. It is straight-forward to compute that the one-sigma contour of a two-dimensional Gaussian marks a 39.4\% confidence region, whereas in three dimensions it is just a 19.9\% confidence region.\footnote{In order to obtain the two-dimensional result, solve the integral $\int_0^{2\pi}d\varphi\int_0^\sigma dr\,r\,\frac{1}{2\pi\sigma^2}\exp\left[-\frac{r^2}{2\sigma^2}\right]$ that assumes a spherically symmetric Gaussian given in polar coordinates.} In general, the confidence $c_P$ contained inside a one-sigma contour of a $P$-dimensional Gaussian with $P>1$ is given by,
\begin{equation}\label{eq:confidence_multivariate_Gaussians}
c_P = \frac{1}{(2\pi)^{P/2}}2\pi\,2^{P-2}\int_0^1 dr\,r^{P-1}e^{r^2/2} \,\textrm{.}
\end{equation}
Table \ref{tab:confidence_multivariate_Gaussians} gives $c_P$ for $P$-dimensional Gaussians with $P\leq 10$, in order to give an impression of how quickly the confidence declines. Evidently, one needs to be very careful when interpreting one-sigma contours in more than one dimension.

\begin{table}
\begin{center}
\begin{tabular}{c|cccccccccc}
\hline
$P$ & 1 & 2 & 3 & 4 & 5 & 6 & 7 & 8 & 9 & 10 \\
\hline
$c_P$ & 0.6827 & 0.3935 & 0.1988 & 0.1149 & 0.0715 & 0.0466 & 0.0314 & 0.0217 & 0.0152 & 0.0109 \\
\hline
\end{tabular}
\end{center}
\caption{Confidence $c_P$ contained within a 1$\sigma$ contour of a $P$-dimensional Gaussian as given by Eq. (\ref{eq:confidence_multivariate_Gaussians}). In ten dimensions, the 1$\sigma$ contour contains a ridiculously small confidence of $\approx 1.1\%$.}
\label{tab:confidence_multivariate_Gaussians}
\end{table}

If the central-limit theorem does not apply -- e.g., because the number $N$ of measured data is small or the likelihood function itself is not well-behaved -- things get even more involved. Nonlinear correlations in the parameters, i.e., ``banana-shaped'' error contours, are an obvious indicator for this case. The symmetric confidence region can still be defined easily, but it obviously lacks the ability to describe parameter correlations. Identifying the ``shortest region'' or ``central region'' may be computationally very expensive. \citet{Barlow1993} recommends a definition of the confidence region via the contour at which the logarithmic likelihood is 0.5 lower than that at its maximum, i.e., where the likelihood function takes $1/e$ of its maximum value. However, the degree of confidence of the resulting region strongly depends on the number of dimensions, similarly to the Gaussian case discussed above.

\begin{wrapfigure}{right}{8cm}
\includegraphics[width=8.0cm]{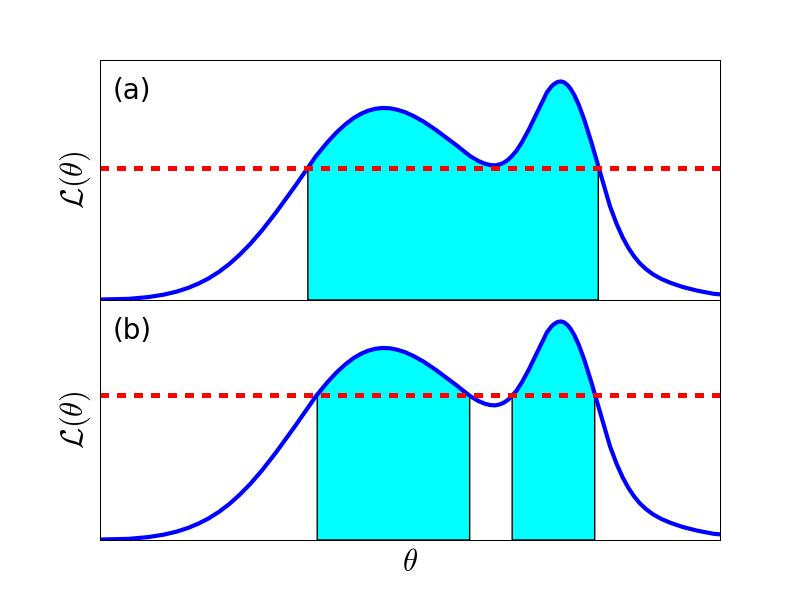}
\caption{Confidence regions for an example likelihood function resulting from Barlow's method (a) and the alternative method presented in this manuscript (b). The horizontal dashed red lines indicate the slices at $\mathcal L_0$. In panel (a) Barlow's region has 80.4\% confidence. In panel (b) the 68.3\% confidence region consists of two regions that are not connected.}
\label{fig:errorEstimateGeneral}
\end{wrapfigure}

As an alternative approach I now discuss a method that is designed to provide a 68.3\% confidence region. Similarly to the method recommended by \citet{Barlow1993}, it employs contours on which the likelihood function is constant. The recipe is as follows:
\begin{enumerate}
\item Locate the maximum of the likelihood function, $\mathcal L_\textrm{max}$, i.e., perform the maximum-likelihood estimation of the parameters.
\item Identify the contours where the likelihood function takes some constant value $\mathcal L_0<\mathcal L_\textrm{max}$ and integrate $\mathcal L$ over these regions in order to get their confidence level.
\item Adjust the contour level $\mathcal L_0$ such that the resulting region in parameter space has 68.3\% confidence.
\end{enumerate}
We visualise this method in Fig. \ref{fig:errorEstimateGeneral} where, for the sake of visualisation, there is only a single parameter. Moreover, Fig. \ref{fig:errorEstimateGeneral} also demonstrates that this method may provide a 68.3\% confidence region that consists of various disconnected regions, if the likelihood function is multimodal. Actually, this is not a real problem because the disconnected regions then also indicate the other local maxima. If the likelihood function is highly multimodal as in Fig. \ref{fig:errorEstimateGeneral} and also in Fig. \ref{fig:confidenceIntervalsGeneral}, this clearly makes more sense than using confidence intervals that are symmetric, central or shortest.

\subsubsection{The worst-case scenario\label{sect:worst_case}}

A final word of caution, before specific methods are discussed: As discussed earlier, parameter estimation via maximising a likelihood function and quantifying its uncertainties via certain intervals is just a makeshift. If it is possible, it greatly simplifies the interpretation because it describes the likelihood function by a simple set of numbers. However, in practice, this procedure may turn out to be infeasible. In such cases, we have to resort to working with the full likelihood function itself, without further simplifications.

\section{Error estimation for model-based parameters}

I now discuss how to perform an error estimation in the case of model-based parameter estimates. In particular, I discuss four different methods -- brute-force grids, varying $\chi^2$, the Fisher matrix, and Monte-Carlo methods -- which are very common in astronomy.

\subsection{Brute-force grids\label{sect:grids}}

\begin{wrapfigure}{right}{8cm}
\includegraphics[width=8cm]{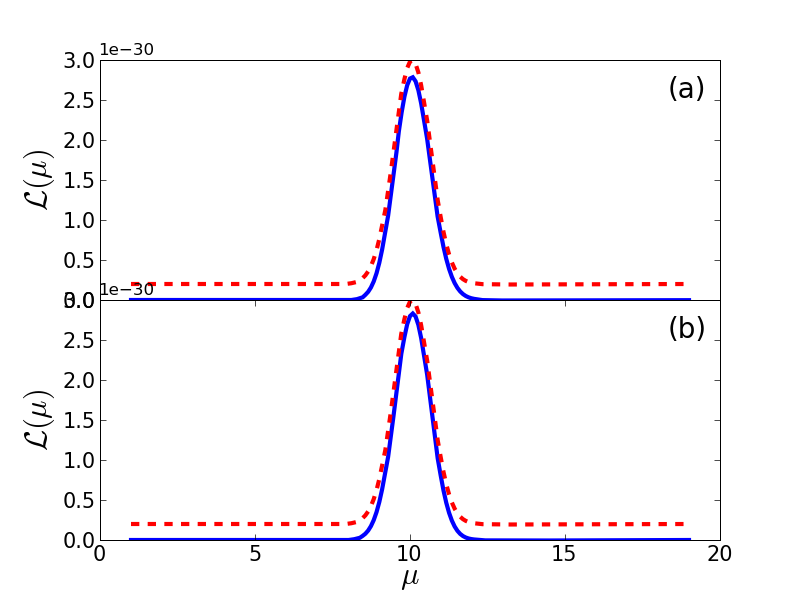}
\caption{Error estimation via brute-force grids for the standard data. Panel (a) assumes the error distribution to be Poissonian, whereas panel (b) assumes it to be Gaussian. Both likelihood functions $\mathcal L(\mu)$ (solid blue curves) peak at $\mu\approx 10.07$. The dashed red curves are matched Gaussians that are slightly shifted upwards for the sake of visibility. The Gaussian's standard deviations (which provide the error estimates) are $\hat\sigma_\mu\approx 0.59$ for panel (a) and $\hat\sigma_\mu\approx 0.59$ for panel (b).}
\label{fig:example_brute_force_grid}
\end{wrapfigure}

A very straight-forward approach to both parameter estimation and subsequent error estimation are brute-force grids. In this case, we span a (rectangular) grid in the parameter space and evaluate the likelihood function at every grid point. If we have only one or two model parameters, we can then plot the likelihood function directly and we can also directly infer the error estimates from the contour lines.

This method is very robust, because it only relies on the assumption that the error distribution of the measured data is correct, i.e., that we are certain about the likelihood function. However, if the number of model parameters is large, this method becomes computationally infeasible. Computational infeasibility may even be an issue for only two or even a single model parameter, if it is very expensive to evaluate the likelihood function.

\subsubsection{Application to example data\label{applic:grids}}

This method is applied to the example data of Table \ref{tab:standard_data_set} and used to try to estimate the mean and its error. For candidate values of $\mu\in [1,19]$ the likelihood function $\mathcal L(\mu)$ is computed assuming the data's error distribution to be Poisson and Gaussian. Figure \ref{fig:example_brute_force_grid} shows $\mathcal L(\mu)$ for both cases. The function $\mathcal L(\mu)$ is then matched by a Gaussian\footnote{This is motivated by the central-limit theorem, but the Gaussian approximation needs to be checked.} and the standard deviation of this Gaussian is the error estimate. This provides the estimate $\hat\sigma_\mu\approx 0.59$ for both Poisson and Gaussian error distributions. Concerning example 3, Figure \ref{fig:binomial_distributions} is essentially a brute-force grid parameter estimation.

\subsection{Varying $\chi^2$\label{sect:varying_chi2}}

$\chi^2$ has already been introduced in Eq. (\ref{eq:def:chi2}). Let us assume that the model parameters were estimated by minimising $\chi^2$. We then vary the model parameters slightly around the optimal values such that $\chi^2$ changes by less than 1. In other words, we look for the contour where $\chi^2=\chi^2_\textrm{min}+1$, thereby defining the error estimate of the model parameters. The basic idea here is that if the likelihood function were Gaussian, this would yield the $1\sigma$ contour.

The crucial assumption of this method is that the error distribution of the measured data is indeed Gaussian, because otherwise using a $\chi^2$ does not make sense (Sect. \ref{sect:example_1}). Moreover, this method relies on an accurate measurement of the data errors $\sigma_n$.

\subsubsection{An example of bad practice}

In this section, I want to discuss an example of bad practise that I discovered as a refereed publication, which provides a couple of lessons to learn. For obvious reasons, I will not give a reference here. The publication pushes the method of varying $\chi^2$ even further in an attempt to ensure that the measurement errors $\sigma_n$ are correct. Fitting a model with $P$ parameters to $N$ data points, the authors demand that
\begin{equation}\label{eq:def:reduced_chi2}
\chi^2 = N-P \quad\Leftrightarrow\quad \chi^2_\textrm{red}=\frac{\chi^2}{N-P}=1 \,\textrm{,}
\end{equation}
where $\chi^2_\textrm{red}$ is called ``reduced $\chi^2$''. Loosely speaking, this means that each degree of freedom contributes one standard deviation. The authors then try to correct for potentially wrong values of the $\sigma_n$ by rescaling them such that $ \chi^2_\textrm{red}=1$.

\begin{wrapfigure}{right}{8cm}
\includegraphics[width=8cm]{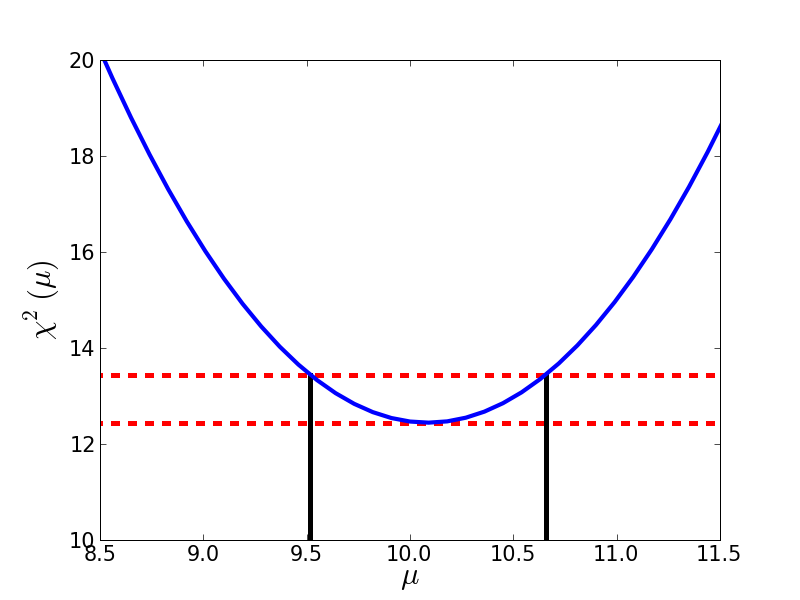}
\caption{Error estimation via varying $\chi^2$ for the standard data. The minimum of $\chi^2$ occurs at $\hat\mu\approx 10.09$ as suggested by Eq. (\ref{eq:ex1_general_estimator}). There $\chi^2\approx 12.4$, as indicated by the lower horizontal, dashed, red line. The upper horizontal, dashed, red line indicates the value $12.4+1.0=13.4$. The error estimates are defined by those points where $\chi^2=13.4$, as indicated by the two vertical solid black lines. The resulting error estimate is $\hat\sigma_\mu\approx 0.57$.}
\label{fig:example_varying_chi2}
\end{wrapfigure}

There are several objections to this method. In order to see this, we need to consider the assumptions involved:
\begin{enumerate}
\item The error distribution has to be Gaussian, as in varying $\chi^2$. The authors do not justify this assumption.
\item The model has to be \textit{linear} in \textit{all} $P$ fit parameters. If the model is nonlinear, we cannot demand that $\chi^2_\textrm{red}=1$, because the derivation of $\chi^2 = N-P$ implicitly assumes linearity in all parameters \citep[e.g.\ cf.][]{Barlow1993}.\footnote{More precisely: We need to know the number of degrees of freedom in order to evaluate $\chi^2_\textrm{red}$. As we are going to show in an upcoming publication, estimating the number of degrees of freedom is possible but nontrivial for linear models. For nonlinear models, however, estimating the number of degrees of freedom is virtually impossible, unless the central-limit theorem provides a good approximation.} Again, the authors do not comment on this, in fact, they even never explicitly say what their model is.
\item By demanding $\chi^2_\textrm{red}=1$, we explicitly claim that the model we are using is the \textit{correct} model that was underlying the data. This is a rather optimistic claim. Of course, we would like our model to be the truth, but this claim requires justification. Moreover, in many practical situations we are actually not interested in the truth, but rather make use of a model which expresses a reasonable simplification.
\end{enumerate}
Even if all these assumptions are met, the method is in fact only applicable if the degrees of freedom $N-P$ are large. The reason is that the uncertainty in the measured data does not only cause an uncertainty in the model parameters, but also an uncertainty in the value of $\chi^2$ itself. The value of $\chi^2$ is subject to the so-called $\chi^2$-distribution \citep[e.g.\ see][]{Barlow1993} whose expectation value is indeed $N-P$. However, this distribution is not sharp but has a nonzero variance of $2(N-P)$. Consequently, if $N-P$ is small, there is a large relative uncertainty on the value of $\chi^2$. This means\ $\chi^2$ may deviate substantially from $N-P$ even though the model is linear and correct.

\subsubsection{Application to example data\label{applic:vary_chi2}}

In Figure \ref{fig:example_varying_chi2}, this method is used to estimate the error on the mean estimator given by Eq. (\ref{eq:ex1_general_estimator}). Given the data of Table \ref{tab:standard_data_set} and $\hat\mu\approx 10.09$, the minimal value of $\chi^2$ is $\approx 12.4$.\footnote{Note that this deviates from $N-1=29$ due to uncertainty in $\chi^2$ induced by the small number of data points.} The next step is to look for those values of $\mu$ where $\chi^2$ takes the values $12.4+1.0=13.4$. In this simple case, these two points are symmetric around the minimum because the model is linear. The resulting error estimate is $\hat\sigma_\mu\approx 0.57$ as in Sect. \ref{applic:grids}. For more general models, the error estimates can be asymmetric. This method is \textit{not} applicable to example 3, because the underlying error distribution is not Gaussian.

\subsection{Fisher matrix\label{sect:fisher_matrix}}

Error estimation via the Fisher matrix is a very popular approach \citep[e.g.\ see][]{Heavens2009}. Let us assume we have fitted our model parameters to our data via maximising the logarithmic likelihood (e.g.\ minimising $\chi^2$ in case of Gaussian noise). This method is based on the central-limit theorem which tells us that any well-behaved likelihood function is \textit{asymptotically} Gaussian near its maximum. We may therefore write
\begin{equation}\label{eq:gaussian_L}
\mathcal L(\theta) \propto \exp\left[-\frac{1}{2}(\vec\theta-\vec\theta_0)^T\cdot\mat\Sigma^{-1}\cdot(\vec\theta-\vec\theta_0)\right] \,\textrm{,}
\end{equation}
which is a $P$-dimensional ($P$-variate) Gaussian with mean $\vec\theta_0$ and covariance matrix $\mat\Sigma$. This covariance matrix is the desired error estimate. On the diagonals it contains the variance estimates of each individual $\theta_p$ and the off-diagonals are the estimates of the covariances \citep[see e.g.][for more information about covariances]{Barlow1993}. Comparing Eqs. (\ref{eq:expanion_L}) and (\ref{eq:gaussian_L}), we identify
\begin{equation}
\hat{\mat\Sigma} = \left(-\frac{\partial^2\log\mathcal L}{\partial\theta_i\partial\theta_j}\right)^{-1}  \,\textrm{.}
\end{equation}
Care should be taken with the order of indices and matrix inversion. The matrix of second derivatives of $\log\mathcal L$ is called ``Fisher matrix''. If the second derivatives of $\log\mathcal L$ can be evaluated analytically, this method may be extremely fast from a computational point of view. However, if this is impossible, they can usually also be evaluated numerically. By construction, this method can only describe elliptical error contours. It is impossible to obtain ``banana-shaped'' error contours from this method.

Of course, this method also invokes assumptions that have to be checked. Those assumptions are:
\begin{enumerate}
\item The error distribution of the measurements is known, i.e., the likelihood function is defined correctly.
\item The second-order Taylor expansion of Eq. (\ref{eq:expanion_L}) is a good approximation.
\end{enumerate}
This second assumption is the actual problem. Although the central-limit theorem ensures this asymptotically, Fig. \ref{fig:failure_fisher} shows an example where this assumption of the Fisher matrix breaks down. There are two simple tests to check the validity of the resulting covariance-matrix candidate $\hat{\mat\Sigma}$. A valid covariance matrix has to be positive definite, i.e., $\vec x^T\cdot\hat{\mat\Sigma}\cdot\vec x>0$ for \textit{any} nonzero vector $\vec x$, and both tests try to check this:
\begin{enumerate}
\item Compute the determinant $\det\hat{\mat\Sigma}$. If $\det\hat{\mat\Sigma}\leq 0$, $\hat{\mat\Sigma}$ is \textit{not} valid.
\item Diagonalise the matrix $\hat{\mat\Sigma}$ in order to determine its eigenvalues. If \textit{any} eigenvalue is negative or zero, $\hat{\mat\Sigma}$ is \textit{not} valid.
\end{enumerate}
The first test is usually easier to perform, whereas the second test is more restrictive. It is strongly recommended that these tests are applied whenever this method is used. Unfortunately, these tests are only rule-out criteria. If $\hat{\mat\Sigma}$ fails any of these tests, it is clearly ruled out. However, if it passes both tests, we still cannot be sure that $\hat{\mat\Sigma}$ is a good approximation, i.e., that the Gaussian is indeed a decent approximation to the likelihood function at its maximum. Nevertheless, the major advantage of this method is that it is very fast and efficient, in particular if we can evaluate the second derivatives of $\log\mathcal L$ analytically.

There are also situations where the Fisher matrix is definitely correct. This is the case for Gaussian measurement errors and linear models. In this case $\mathcal L(\vec\theta)$ is truly a Gaussian even without any approximation. For instances, inspect Eq. (\ref{eq:def:chi2}): $\chi^2$ is a quadratic function of $\mu$, and since $\mathcal L(\mu)\propto e^{-\chi^2/2}$ the likelihood function is a Gaussian.

\subsubsection{Example 1 revisited}

In order to see the Fisher matrix ``in action'', we now return to example 1 from Sect. \ref{sect:example_1}. The task was to estimate the mean $\mu$ of $N$ data points $\{x_1,x_2,\ldots,x_N\}$ that are drawn from a Gaussian error distribution. The maximum-likelihood estimator for $\mu$ is given by Eq. (\ref{eq:one}). The task now is to use the Fisher matrix in order to derive the error estimate for Eq. (\ref{eq:one}). Equation (\ref{eq:first_derivative}) is derived once more with respect to $\mu$,
\begin{equation}\label{eq:second_derivative}
\frac{d^2\log\mathcal L(D;\mu)}{d\mu^2} = \sum_{n=1}^N\frac{1}{\sigma_n^2}  \,\textrm{.}
\end{equation}
For the error estimate of $\hat\mu$ this then yields
\begin{equation}\label{eq:example_Fisher_Gauss}
\hat\sigma_\mu^2 = \left(\sum_{n=1}^N\frac{1}{\sigma_n^2}\right)^{-1} \,\textrm{.}
\end{equation}
If all $N$ data points are again assumed to have identical errors, $\sigma_n=\sigma$, this simplifies to $\hat\sigma_\mu^2 = \sigma^2/N$ which is the error estimate for the arithmetic mean in case of Gaussian measurement errors.

\subsubsection{Example 2 revisited}

The same exercise is repeated for example 2, where the error distribution is assumed to be Poissonian. The second derivative of Eq. (\ref{eq:ex2_first_derivative}) with respect to $\mu$ reads
\begin{equation}
\frac{d^2\log\mathcal L(D;\mu)}{d\mu^2} = -\frac{1}{\mu^2}\sum_{n=1}^N x_n  \,\textrm{.}
\end{equation}
For the error estimate of $\hat\mu$ this yields
\begin{equation}\label{eq:example_Fisher_Poisson}
\hat\sigma_\mu^2 = \frac{\hat\mu^2}{\sum_{n=1}^N x_n} = \frac{\hat\mu}{N} \,\textrm{,}
\end{equation}
where $\sum_{n=1}^N x_n = N\hat\mu$ has been identified according to the estimator of Eq. (\ref{eq:poisson_estimator}).

\subsubsection{Application to example data\label{applic:fisher}}

Inserting the example data of Table \ref{tab:standard_data_set} into Eq. (\ref{eq:example_Fisher_Gauss}), an error estimate of $\hat\sigma_\mu\approx 0.57$ is obtained, which is in agreement with the error estimates from Sections \ref{applic:grids} and \ref{applic:vary_chi2}. In the case of a Poisson error distribution, insertion of the example data of Table \ref{tab:standard_data_set} into Eq. (\ref{eq:example_Fisher_Poisson}) yields $\hat\sigma_\mu\approx 0.58$, which is also in agreement with the result of Sect. \ref{applic:grids}. Both results are reliable, since in both cases the likelihood function is well approximated by a Gaussian, as we have seen in Fig. \ref{fig:example_brute_force_grid}.

\subsubsection{Example 3 revisited}

Let us compute the second derivative of the logarithmic likelihood given by Eq.\ (\ref{eq:binomial_loglik}) w.r.t.\ $f$. We obtain,
\begin{equation}
\frac{d^2\log\mathcal L}{d f^2} = -\frac{n}{f^2}  -\frac{N-n}{(1-f)^2} = -\frac{N^3}{n(N-n)} \;\textrm{,}
\end{equation}
where we have inserted the maximum-likelihood estimator $\hat f=n/N$. Hence, the error estimate pretends to be
\begin{displaymath}
\sigma_f^2 = \frac{n(N-n)}{N^3} \;\textrm{,}
\end{displaymath}
which yields $\sigma_f^2=0$ in example 3.1 and $\sigma_f^2\approx 0.004$ in example 3.2. Now we are lucky, because $\sigma_f^2=0$ tells us that we forgot that in example 3 the likelihood function is binomial and not Gaussian, i.e., this whole calculation was \textit{nonsense}. Unfortunately, this is not necessarily that obvious, as the result for example 3.2 shows.

\subsection{Monte-Carlo methods\label{sect:MC_methods}}

Monte-Carlo methods directly draw samples from the likelihood function, i.e., they guess values of the fit parameters and accept them with the probability defined by the corresponding value of the likelihood function. Although these methods may appear difficult at first glance, Monte-Carlo sampling is actually the most intuitive approach to error estimation. The reason is its similarity to measuring errors of data by repeating the measurement process and monitoring the results. The strength of all Monte-Carlo methods is that they use a minimum amount of assumptions. Their sole and only assumption is that the error distribution of the measured data is known correctly, i.e., that we are certain about the likelihood function. There are no further requirements such as Gaussianity, which renders this approach very general.

There are many different types of Monte-Carlo methods for many different situations. It would be beyond the scope of this manuscript to explain these methods. Instead, I explain which methods are useful in which kinds of situations and refer the interested reader to the literature. \citet{MacKay2003}, for example, provides an excellent introduction to all the methods named here.

\begin{wrapfigure}{right}{8cm}
\includegraphics[width=8cm]{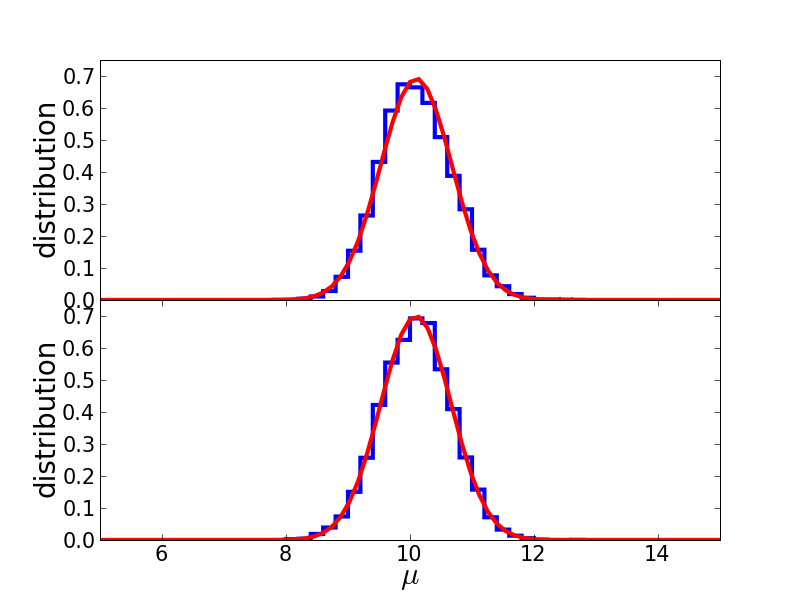}
\caption{Error estimation via Monte-Carlo sampling for the standard data. The distributions of $\mu$ resulting from the MCMC algorithm assuming Poisson errors in panel (a) and Gaussian errors in panel (b) are both well approximated by a Gaussian. In panel (a) the Gaussian is given by $\hat\mu=10.10$ and $\hat\sigma_\mu=0.58$. In panel (b) the Gaussian is given by $\hat\mu=10.10$ and $\hat\sigma_\mu=0.57$.}
\label{fig:example_MCMC}
\end{wrapfigure}

The first criterion in choosing a particular Monte-Carlo algorithm is the number of model parameters. If the number of parameters is small -- say 1 or 2 -- we can use one of the following Monte-Carlo algorithms: uniform sampling, importance sampling, or rejection sampling. However, if the number of parameters is large, these algorithms quickly become inefficient, i.e., computationally expensive. In the case of many model parameters, there is the family of Markov-chain Monte-Carlo (MCMC) algorithms, whose computation times scale linearly with the number of model parameters. The most famous (and most simple) type of MCMC algorithm is the Metropolis-Hastings algorithm. However, this algorithm involves a stepsize for each parameter that needs to be fine tuned by hand. This is a real problem in the case of many parameters. We therefore recommend the slice-sampling algorithm, which also involves a stepsize but it is tuned automatically without human interaction. In fact, there is also Gibbs sampling which does not involve any stepsizes at all. However, Gibbs sampling can only be used directly for trivial models. In order to employ Gibbs sampling in a wider context, one has to combine it with other one-dimensional sampling algorithms.

In practice, we start an MCMC algorithm at the estimated maximum $\vec\theta_0$ of the likelihood function and let it draw $M$ samples of model parameters $\{\vec\theta_1,\ldots,\vec\theta_M\}$ from the likelihood function.\footnote{Initialising the MCMC at the optimum $\vec\theta_0$ also has the advantage that we do not need to use any convergence diagnostics, since we are already at the optimum. Convergence diagnostics for MCMC algorithms are still an open issue \citep{Cowles1996}.} All we then need to do is to analyse the distribution of this set of parameters.\footnote{In practice we often need to ``thin'' this sequence due to autocorrelations in the Markov chain, e.g., by selecting every tenth point and discarding all others.} For instance, assuming Gaussianity, we can do this via estimating the mean and the covariance of this sample. However, the distribution can also be analysed more generally such that it is possible to describe nonlinear dependencies between different model parameters, e.g., ``banana shaped'' error contours.

\subsubsection{Application to example data\label{applic:MCMC}}

Once again, this method is applied to the example data given in Table \ref{tab:standard_data_set}. A Metropolis-Hastings algorithm is used with stepsizes inspired by the error estimates obtained previously. For likelihood functions assuming Poissonian and Gaussian error distributions, respectively, the MCMC algorithm is iterated 100,000 times. Afterwards, the resulting sequences are thinned by picking out every tenth data point and discarding everything else. From the remaining 10,000 data points the arithmetic mean and its standard deviation are estimated. In Fig.\ \ref{fig:example_MCMC}, the resulting data distribution is overplotted by a corresponding Gaussian, which provides the error estimate. For the Poisson error distribution $\hat\sigma_\mu=0.58$ is obtained, and $\hat\sigma_\mu=0.57$ for Gaussian error distribution. This is again in agreement with previous results.

\section{Error estimation for model-independent parameters}

So far we have focused our attention on error estimation for model-based parameters. Model-based inference usually involves an optimisation problem which can complicate matters. Therefore, model-independent parameters that usually do \textit{not} involve any optimisation are very popular in astronomy. I now explain how to estimate errors for parameters of this type.

\subsection{Error propagation\label{sect:error_propagation}}

One simple method is error propagation. If we can express the parameter as a function of the measured data and if the error distribution of the data is Gaussian, we can employ Gaussian error propagation.

To give an example, let us consider the flux $F$ of a galaxy. If the image background has been subtracted correctly, the flux is just given by the sum of all $N$ pixel values $f_i$ in the image,
\begin{equation}\label{eq:flux_estimator}
\hat F=\sum_{i=1}^N f_i \,\textrm{.}
\end{equation}
As argued in Sect. \ref{sect:data_errors}, the error distribution in photometric images is in excellent approximation to Gaussian, if the exposure time was long enough. If we denote the measurement error of pixel $i$ by $\sigma_i$, we can then estimate the error of $\hat F$ via
\begin{equation}
\hat\sigma_F^2 = \sum_{i=1}^N \left(\sigma_i\frac{\partial \hat F}{\partial f_i}\right)^2 = \sum_{i=1}^N \sigma_i^2 \,\textrm{,}
\end{equation}
which is fairly simple in this case. However, this can become very cumbersome for more general model-independent parameters. In particular, it is impossible if a model-independent parameter involves an operation on the measured data that is not differentiable, e.g., selecting certain data points. In fact, error propagation can also be applied to model-based parameter estimates, if  these estimators can be expressed as differentiable functions of the data, e.g., as is the case for linear models with Gaussian measurement errors. For instance, Equation (\ref{eq:example_Fisher_Gauss}) can also be derived from Eq. (\ref{eq:ex1_general_estimator}) using this method, giving the same result for the example data of Table \ref{tab:standard_data_set}. However, in general, this is not the case.

\subsection{Resampling the data\label{sect:mcmc_resampling}}

A very elegant method for error estimation is to \textit{resample} the measured data \citep[for an example see e.g.][]{Burtscher2009}. Again, the assumption is that we know the correct error distribution of the measured data. For the sake of simplicity, let us assume this error distribution is Gaussian. For each data point $x_n$ we invoke a Gaussian with mean $x_n$ and standard deviation $\sigma_n$ as measured during the experiment. We can now sample a new data point $x_n^\prime$ from this distribution.\footnote{In fact, this method is a Monte-Carlo method, too.} Doing this for all pixels, we get an ``alternative'' noise realisation of the measurement, e.g., a new image that can be interpreted as an alternative measurement result. We then estimate our model-independent parameter for this resampled data and the result will differ slightly from that obtained from the actual data. Repeating this resampling process, e.g., 100 times, and monitoring the resulting parameter estimates, we get a distribution of this model-independent parameter from which we can then infer an \textit{upper limit} for the uncertainty. For instance, if the resulting parameter distribution is approximately Gaussian, the upper limit will be given by the Gaussian's standard deviation.

Why does this method only provide an upper limit for the uncertainty? The reason is that even though we are using the correct error distribution of the measured data, we are centering this error distribution at the measured value instead of the (unknown) true value. This introduces additional scatter and leads us to overestimate the uncertainty. Nevertheless, it may still be acceptable to use the overestimated uncertainty as a conservative estimate, depending on the precise scientific question.

This method is a very intuitive approach to error estimation, because it \textit{simulates} repeated measurements of the data. It can also be applied to error estimation for model-based parameter estimates.

\subsubsection{Application to example data\label{applic:resampling}}

\begin{figure}
  \begin{minipage}[t]{0.45\textwidth}\centering
\includegraphics[width=8cm]{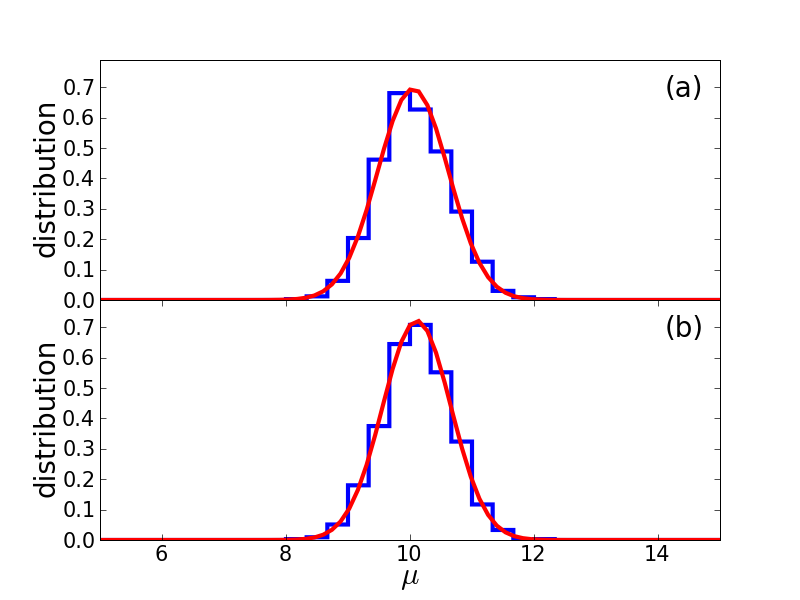}
\caption{Error estimation via resampling the data, using the standard data. The distributions of $\mu$ resulting from the resampling procedure assuming Poisson errors in panel (a) and Gaussian errors in panel (b) are both well approximated by a Gaussian. In panel (a) the Gaussian is given by $\hat\mu=10.05$ and $\hat\sigma_\mu=0.57$. In panel (b) the Gaussian is given by $\hat\mu=10.11$ and $\hat\sigma_\mu=0.55$.}
\label{fig:example_resampling_data}
  \end{minipage}
  \hspace*{0.05\textwidth}
  \begin{minipage}[t]{0.45\textwidth}\centering
\includegraphics[width=8cm]{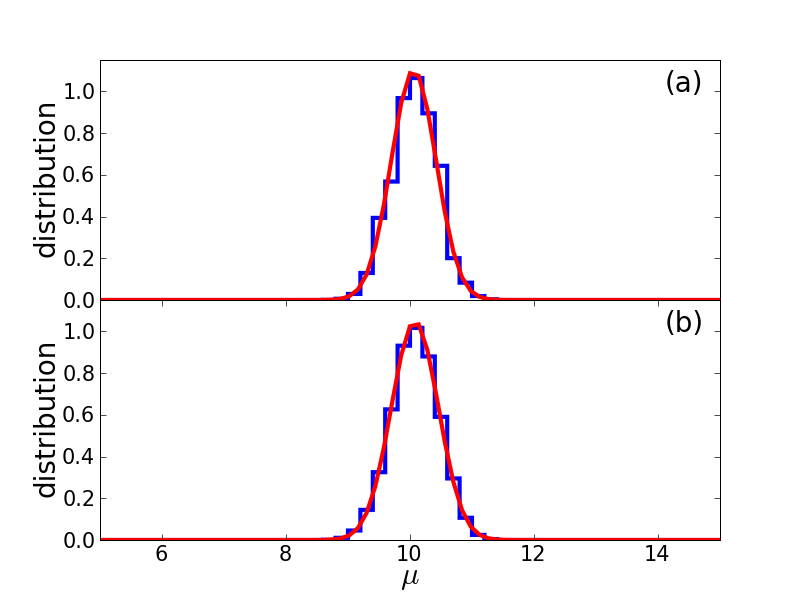}
\caption{Error estimation via bootstrapping the data, using the standard data. The distributions of $\mu$ result from the bootstrapping procedure and estimating the Poisson mean via Eq. (\ref{eq:poisson_estimator}) in panel (a) and estimating the Gaussian mean via Eq. (\ref{eq:ex1_general_estimator}) in panel (b). Both distributions are well approximated by a Gaussian. In panel (a) the Gaussian is given by $\hat\mu=10.06$ and $\hat\sigma_\mu=0.36$. In panel (b) the Gaussian is given by $\hat\mu=10.08$ and $\hat\sigma_\mu=0.38$.}
\label{fig:example_bootstrapping}
  \end{minipage}
\end{figure}

Again, this method is applied to the example data of Table \ref{tab:standard_data_set}. The measured data points are assumed to be the means of the error distribution (Poisson and Gaussian) and each data point is resampled from its error distribution. For this resampled data, the Poisson mean is estimated via Eq. (\ref{eq:poisson_estimator}) and the Gaussian mean via Eq. (\ref{eq:ex1_general_estimator}). This is repeated 1,000 times and the estimated means are monitored. Figure \ref{fig:example_resampling_data} shows the resulting distribution of mean values. The error estimates agree well with those obtained from the other methods.

\subsection{Bootstrapping\label{sect:bootstrapping}}

Bootstrapping \citep{Efron1979,Hastie2009} is another resampling method for error estimation that can be applied to model-based as well as model-independent parameter estimation. Let us assume we have $N$ measurements $\{x_1,x_2,\ldots,x_N\}$ from which we estimate some parameter. Again, we resample our data in order to create ``alternative'' data sets from which we then repeatedly estimate the parameter of interest, monitoring its distribution as before. However, the details of the resampling process are different from those in Sect. \ref{sect:mcmc_resampling}. Instead of resampling each data point from its individual error distribution, we draw new samples from the measured data set itself. Drawing these ``bootstrap'' samples is done with replacement, i.e., the same data point can occur multiple times in our bootstrap sample. To give an example, we consider a data set $\{x_1,x_2,x_3,x_4\}$. Some examples for its bootstrap samples are:
\begin{itemize}
\item $\{x_1,x_2,x_3,x_4\}$ itself, of course.
\item $\{x_1,x_2,x_1,x_4\}$,
\item $\{x_1,x_2,x_2,x_4\}$,
\item $\{x_1,x_3,x_3,x_3\}$,
\item $\{x_2,x_2,x_2,x_2\}$, which is bad but possible.
\end{itemize}
As the same data point can occur multiple times but ordering is not important, for $N$ data points the total number of possible different bootstrap samples is
\begin{equation}
\left(\begin{array}{c}
2N-1 \\ N
\end{array}\right) = \frac{(2N-1)!}{N!(N-1)!} \,\textrm{.}
\end{equation}
In practice, the number of bootstrap samples chosen is set to some useful number, where ``useful'' is determined by the trade-off between computational effort and the desire to have as many samples as possible in order to get a good estimate.

The major advantage of bootstrapping is that the error distribution of the measured data does \textit{not} need to be known (unless the parameter estimation itself requires it). The crucial assumption here is that the measured data sample itself encodes the information about its error distribution. However, the parameter estimation must be capable of dealing with the bootstrapped samples which, in general, include certain data points multiple times while completely lacking other data points. For instance, the flux estimator of Eq. (\ref{eq:flux_estimator}) would \textit{not} be capable of handling bootstrap samples, since \textit{all} pixels have to contribute precisely \textit{once}. Nevertheless, if we know the data's error distribution, we should really exploit this knowledge by using, e.g., resampling instead of bootstrapping.

\subsubsection{Application to example data\label{applic:bootstrapping}}

Finally, also bootstrapping is applied to the data of Table \ref{tab:standard_data_set}. This data sample contains $N=30$ values, i.e., the total number of possible bootstrap samples is $\approx 5.9\cdot 10^{16}$. Then, 10,000 bootstrap samples are drawn and for every sample the Poisson mean is estimated via Eq. (\ref{eq:poisson_estimator}) and the Gaussian mean via Eq. (\ref{eq:ex1_general_estimator}). Figure \ref{fig:example_bootstrapping} shows the resulting distributions. Obviously, the mean values are estimated correctly. However, the errors are underestimated, especially in the case of the Gaussian. The likely explanation is that the data sample is not large enough in order to contain sufficient information about its underlying error distribution.

\section{Propagating measurement errors through data-reduction pipelines\label{sec:propagate_pipelines}}

Usually, directly measured data has to be preprocessed before it can be analysed further. The preprocessing is typically done by some data-reduction pipeline. For instances, the data coming out of a spectrograph is in its raw state rather useless. Before one can analyse it, one has to subtract the bias, correct for the flat field response of the CCD, and calibrate wavelength and flux. This preprocessing is usually done either using complete instrument-specific pipelines, more general software routines such as those found in IRAF or MIDAS, or a combination of both. Now the question arises, how to propagate the errors on the initial measurements through such complex preprocessing?

In general, an analytic error propagation such as that discussed in Sect. \ref{sect:error_propagation} is impossible, either because the data-reduction pipeline is too complex, or because the pipeline is used as a ``black box''. Nevertheless, it is possible to propagate the errors through the pipeline via resampling (Sect. \ref{sect:mcmc_resampling}), though it may be computationally expensive. Let us outline this method for our previous example of spectral measurements. We do have the raw spectral data, the measured bias fields and flat fields. We resample each of these fields as described in Sect. \ref{sect:mcmc_resampling}, say $N$ resamplings, assuming the measurement errors are Gaussian (or Poisson, if we count only few photons). Then, we feed each resampling instance through the pipeline and monitor the outcome. The result will be a set of reduced spectra, which provide an estimate of the reduced spectrum's error distribution.

Of course, this method may be computationally expensive in practice.\footnote{Although after the necessary groundwork has been covered, the batch processing of simple spectra usually takes no more than a few seconds a piece.} However, as we have argued earlier, an error estimate is inevitably necessary. Therefore, if this method is the only possibility to get such an error estimate, computational cost is not an argument.\footnote{In fact, one may argue that computational cost is not an argument anyway. If something is computationally too expensive, then one should buy more computers! Unfortunately, this approach is usually hampered by the fact that licensed software is very popular in astronomy (e.g. IDL).}

\section{Summary}

I have discussed different methods for error estimation that apply to model-based as well as model-independent parameter estimates. The methods have been briefly outlined and their assumptions have been made explicit. Whenever employing one of these methods, all assumptions should be checked. Table \ref{tab:overview_error_estimation_methods} summarises all the methods discussed here and provides a brief overview of their applicability. It was beyond the scope of this manuscript to describe all methods in detail. Where possible I pointed to the literature for examples. Furthermore, I have also outlined how one can propagate errors through data-reduction pipelines.

My recommendations for error estimation are to use Monte-Carlo methods in case of model-based parameter estimates and Monte-Carlo-like resampling of the measured data in case of model-independent parameter estimates. These methods only require knowledge of the measurement errors but do not invoke further assumptions such as Gaussianity of the likelihood function near its maximum. Bootstrapping may also be an option if sufficient data are available.

I conclude with some recommendations for further reading:
\begin{itemize}
\item \citet{Barlow1993}: An easy-to-read introduction into the basics of statistics without going too much into depth. Recommendable to get a first idea about parameter estimation, error estimation, and the interpretation of uncertainties.
\item \citet{Press2002}: This book contains some very useful chapters about statistical theory, e.g., linear least-squares fitting. It is excellent for looking up how a certain method works, but it is not meant as an introduction to statistics.
\item \citet{Hastie2009}: A good textbook giving a profound introduction into analysing data. However, the main focus of this book is on classification problems, rather than regression problems. Given the rather mathematical notation and compactness, this book requires a certain level of prior knowledge.
\item \citet{MacKay2003}: The focus of this textbook is again mainly on classification, but it also provides a broader overview of concepts of data analysis and contains an excellent introduction to Monte-Carlo methods. This textbook also requires a certain level of prior knowledge.
\end{itemize}

\begin{table}
\begin{center}
\begin{tabular}{lccccc}
\hline\hline
method & section & model-based & model-independent & data error & contours\\
\hline
brute-force grids   & \ref{sect:grids}             & yes & no  & known  & arbitrary \\
varying $\chi^2$    & \ref{sect:varying_chi2}      & yes & no  & Gaussian  & arbitrary \\
Fisher matrix       & \ref{sect:fisher_matrix}     & yes & no  & known  & elliptical \\
Monte-Carlo methods & \ref{sect:MC_methods}        & yes & no  & known  & arbitrary \\
error propagation   & \ref{sect:error_propagation} & depends & yes & Gaussian   & elliptical \\
resampling the data & \ref{sect:mcmc_resampling}   & yes & yes & known  & arbitrary \\
bootstrapping       & \ref{sect:bootstrapping}     & yes & yes & unknown & arbitrary \\
\hline
\end{tabular}
\end{center}
\caption{Methods for error estimation discussed in this manuscript. This table gives a brief overview of each method: Specifically, whether a certain method is applicable to model-based and/or model-independent parameter estimates, whether knowledge about the data's error distribution is necessary, and what kind of error contours can be estimated.}
\label{tab:overview_error_estimation_methods}
\end{table}

\paragraph*{Acknowledgements}

I would like to thank David W. Hogg for reading this manuscript and providing valuable comments on the technical issues. Furthermore, I would like to thank Ellen Andrae and Katherine Inskip for helping me to get this rather technical subject down to something that is hopefully readable. Katherine also helped me to get the examples with spectroscopy right. Last but not least, I want to thank Ada Nebot Gomez-Mor\'an, who complained that she was dearly missing a guide to error estimation during her PhD. I hope she approves of this manuscript. This is the first revised version of the original manuscript.

\bibliographystyle{aa}

\def\physrep{Phys. Rep.}%
\def\apjs{ApJS}%
\def\apj{ApJ}%
\def\apjl{ApJL}%
\def\aj{AJ}%
\def\aap{A\&A}%
\def\aaps{A\&AS}%
\def\mnras{MNRAS}%
        
\bibliography{bibliography}

\end{document}